\documentclass[useAMS,usenatbib]{mn2e}
\usepackage{graphicx}
\usepackage[fleqn]{amsmath}
\usepackage{amssymb}
\usepackage{multirow}

\def\ie{i.e.~}
\def\eg{e.g.~}
\def\HST{{\it{HST}}}
\def\bHST{{\it{(HST)}}}
\def\water{H$_2$O}
\def\methane{CH$_4$}

\def\s{$\sim$}

\title[WFC3 transmission spectroscopy of HD 189733b]{Probing the haze in the atmosphere of HD 189733b with HST/WFC3 transmission spectroscopy}
\author[N. P. Gibson et al.]{
N. P. Gibson$^{1}$\thanks{E-mail: Neale.Gibson@astro.ox.ac.uk},
S. Aigrain$^{1}$,
F. Pont$^{2}$,
D. Sing$^{2}$,
J.-M. D\'{e}sert$^{3}$,
T. M. Evans$^{1}$, \newauthor
G. Henry$^{4}$,
N. Husnoo$^{2}$
and
H. Knutson$^{5}$
\smallskip
\\
$^{1}$Department of Physics, University of Oxford, Denys Wilkinson Building, Keble Road, Oxford OX1 3RH, UK\\
$^{2}$School of Physics, University of Exeter, Exeter, EX4 4QL, UK\\
$^{3}$Harvard-Smithsonian Center for Astrophysics, Cambridge, MA 02138, USA\\
$^{4}$Tennessee State University, 3500 John A. Merritt Blvd., PO Box 9501, Nashville, TN 37209, USA\\
$^{5}$Division of Geological and Planetary Sciences, Caltech, 1200 E California Blvd, Pasadena CA 91125, USA
}

\begin{document}

\date{Accepted 2012 January 30. Received 2012 January 27; in original form 2012 January 12}

\pagerange{\pageref{firstpage}--\pageref{lastpage}} \pubyear{2002}

\maketitle

\label{firstpage}

\begin{abstract}

We present {\it Hubble Space Telescope} near-infrared transmission spectroscopy of the transiting exoplanet HD 189733b, using Wide Field Camera 3. This consists of time-series spectra of two transits, used to measure the wavelength dependence of the planetary radius. These observations aim to test whether the Rayleigh scattering haze detected at optical wavelengths extends into the near-infrared, or if it becomes transparent leaving molecular features to dominate the transmission spectrum. Due to saturation and non-linearity affecting the brightest (central) pixels of the spectrum, light curves were extracted from the blue and red ends of the spectra only, corresponding to wavelength ranges of $1.099-1.168$\,\micron\ and $1.521-1.693$\,\micron, respectively, for the first visit, and $1.082-1.128$\,\micron\ and $1.514-1.671$\,\micron\ for the second. The light curves were fitted using a Gaussian process model to account for instrumental systematics whilst simultaneously fitting for the transit parameters. This gives values of the planet-to-star radius ratio for the blue and red light curves of $0.15650\pm0.00048$ and $0.15634\pm0.00032$, respectively, for visit one and $0.15716\pm0.00078$ and $0.15630\pm0.00037$ for visit 2 (using a quadratic limb darkening law). The planet-to-star radius ratios measured in both visits are consistent, and we see no evidence for the drop in absorption expected if the haze that is observed in the optical becomes transparent in the infrared. This tentatively suggests that the haze dominates the transmission spectrum of HD 189733b into near-infrared wavelengths, although more robust observations are required to provide conclusive evidence.

\end{abstract}


\section{Introduction}

The study of transiting planets is revolutionising our understanding of planets beyond our Solar System. Transiting planets are those which pass between us and their host stars, periodically blocking starlight and producing a characteristic light curve. These are vital to studies of planetary structure, as the radius of a planet and its orbital inclination can be measured from the transit, and coupled with radial velocity data the mass, bulk density and structure may be inferred.

Transiting planets also provide an opportunity to probe atmospheric structure and composition, through transmission and emission (eclipse) spectroscopy. Transmission spectroscopy is a measurement of the planet-to-star radius ratio as a function of wavelength. As the optical depth in the atmosphere is wavelength dependent, so too is the altitude at which the planet becomes opaque to starlight. Wavelength dependent measurements of the planet-to-star radius ratio are therefore sensitive to atomic and molecular species in the atmosphere \citep[\eg][]{Seager_2000,Brown_2001}.

Observations with the {\it Hubble Space Telescope} {\bHST} and {\it Spitzer Space Telescope} have provided the most detailed observations of exoplanet atmospheres to date \citep[see e.g.][]{Charbonneau_2002,Vidal-Madjar_2003,Charbonneau_2005,Deming_2005}. In particular, HD 189733b is one of the most studied exoplanets, being one of the brightest known transiting systems. At optical wavelengths its transmission spectrum is largely featureless, dominated by Rayleigh scattering from a high altitude haze layer from \s0.3--1.0\,\micron\, first detected by \citet{Pont_2008} using the ACS instrument, and recently confirmed using STIS \citep{Sing_2011}. The haze is thought to consist of condensate particles; the most likely candidate being MgSiO$_3$ grains \citep{Lecavelier_2008a,Lecavelier_2008b}. However, Na has been detected using higher resolution observations \citep{Redfield_2008, Huitson_2011}. Transits at UV wavelengths have shown the planet to have an escaping hydrogen atmosphere \citep{Lecavelier_2010}. The situation in the near-infrared is more debated; whilst the transmission spectrum obtained with NICMOS from 1.4--2.5\,\micron\ apparently showed significant \water\ and \methane\ features \citep{Swain_2008}, narrow band photometry with NICMOS \citep{Sing_2009} proved to be inconsistent with this yet consistent with extrapolation of the optical haze. Spitzer observations at 3.6\micron\ are again consistent with the haze, although longer wavelength observations suggest the presence of molecular absorption \citep{Desert_2009,Desert_2011}.

The interpretation of molecular features in the NICMOS transmission spectroscopy requires the haze to become largely transparent somewhere between \s1--1.4\,\micron, and the transmission needs to show significant temporal variation for the spectroscopic and narrow-band observations to be consistent. A more likely explanation is that the systematics in the NICMOS transmission spectroscopy dataset were not fully accounted for as suggested by \citet{Sing_2009}. In \citet{Gibson_2011} we reanalysed the NICMOS transmission spectroscopy, and claimed that standard methods to model instrumental systematic based on linear basis functions were not reliable in this case, and that the detections of molecular species relied on unjustified assumptions about the form of the systematics model. Therefore the uncertainties are largely underestimated and not precise enough to probe for the presence of molecular species. This interpretation was largely confirmed in \citet{Gibson_2011b}, where we developed a Gaussian process (GP) model to extract the transmission spectrum robustly in the presence of such systematics, although it is important to note that this view is still disputed by the original authors \citep{Swain_2011}.

GPs allow the use of non-parametric models for the instrumental systematics, and infers the transit parameters in a fully Bayesian framework. Therefore fewer assumptions are made about the form of the systematics model, providing more robust transmission spectra. The NICMOS transmission spectrum of HD 189733b inferred using GPs does not exclude the presence of molecular species beyond \s1.5\micron; however, if they are present, we cannot constrain them very well with current data. Either way, the question remains of how far the haze observed in the optical extends into the infrared. Observations bridging the gap between the optical and NICMOS data should help resolve the situation.

Here we present observations designed for that purpose; \HST\ transmission spectroscopy of HD 189733 using the Wide-Field Camera 3 (WFC3) from $1.08-1.69$\,\micron. These observations were originally planned with NICMOS as part of program GP 11740 (P.I. F. Pont), before the failure of the NICMOS cooling system to restart. WFC3 was recently used by \citet{Berta_2011} to observe the transmission spectrum of GJ1214b, resulting in near-photon limited observations, and showing the WFC3 will prove a powerful instrument for such measurements. However, here we are using the WFC3 in a noticeably different regime to observe a much brighter star. Sect.~\ref{sect:observations} describes the observations and data reduction, and Sect.~\ref{sect:analysis} describes the analysis of the light curves. Finally Sects.~\ref{sect:results} and \ref{sect:conclusions} present our results and conclusions.

\section{WFC3 observations and data reduction}
\label{sect:observations}

Two transits of HD 189733b were observed with the IR channel of \HST/WFC3 on 2010 September 4 and 2010 November 10, as part of program GP 11740 (P.I. F. Pont). Observations used the G141 grism, and we obtained slitless spectra from $\sim1.08-1.69$\,\micron~with a resolution of $\lambda/\Delta\lambda\sim130$ at the central wavelength. For each visit, 592 spectra were taken, preceded by an image taken with the F167N filter in order to set the wavelength calibration. As HD 189733 is not in the continuous viewing zone of \HST, each transit was observed over 4 half-orbits ($\sim$48 minute blocks). For both visits, the first, second and fourth orbits cover the out-of-transit part of the light curve, and consist of 142, 150 and 150 spectra, respectively, and the third orbit covers the in-transit part of the observation, consisting of 150 spectra.

As HD 189733 is a particularly bright star ($V\sim7.67, K\sim5.54$), we used exposure times of only 0.225 seconds\footnote{It is not possible to defocus WFC3, and at the time of phase II preparation, `driftscan' mode \citep{McCullough_2011} was not available. This involves slowly moving the telescope during exposures to smear out the cross-dispersion PSF, therefore avoiding saturation of the brightest pixels and allowing longer exposures.}. These were taken in {\sc multiaccum} mode with 2 non-destructive reads ({\sc rapid, nsamp=2}). This means that the detector is read out three times per exposure, once at the start, once mid-way, and once at the end of the exposure. Hereafter we refer to these images as the zeroth, first and second readouts. The overheads for bright targets with WFC3 are dominated by reading out the detector, and dumping the temporary storage buffer of WFC3 to the \HST\ solid sate recorder. Therefore the camera was used in {\sc subarray} mode, reading out only the central 128$\times$128 pixels ({\sc irsub128}), which contains most (but not all) of the first order spectrum, but excludes the zeroth order image and the second order spectrum. This reduced the readout time and increased the interval between buffer dumps, eliminating the need for a long buffer dump during each orbit. We used the shortest possible exposure time in this readout mode. This resulted in a cadence of $\sim$18 seconds, the minimum allowed by the WFC3 electronics.

\begin{figure}
\includegraphics[width=85mm]{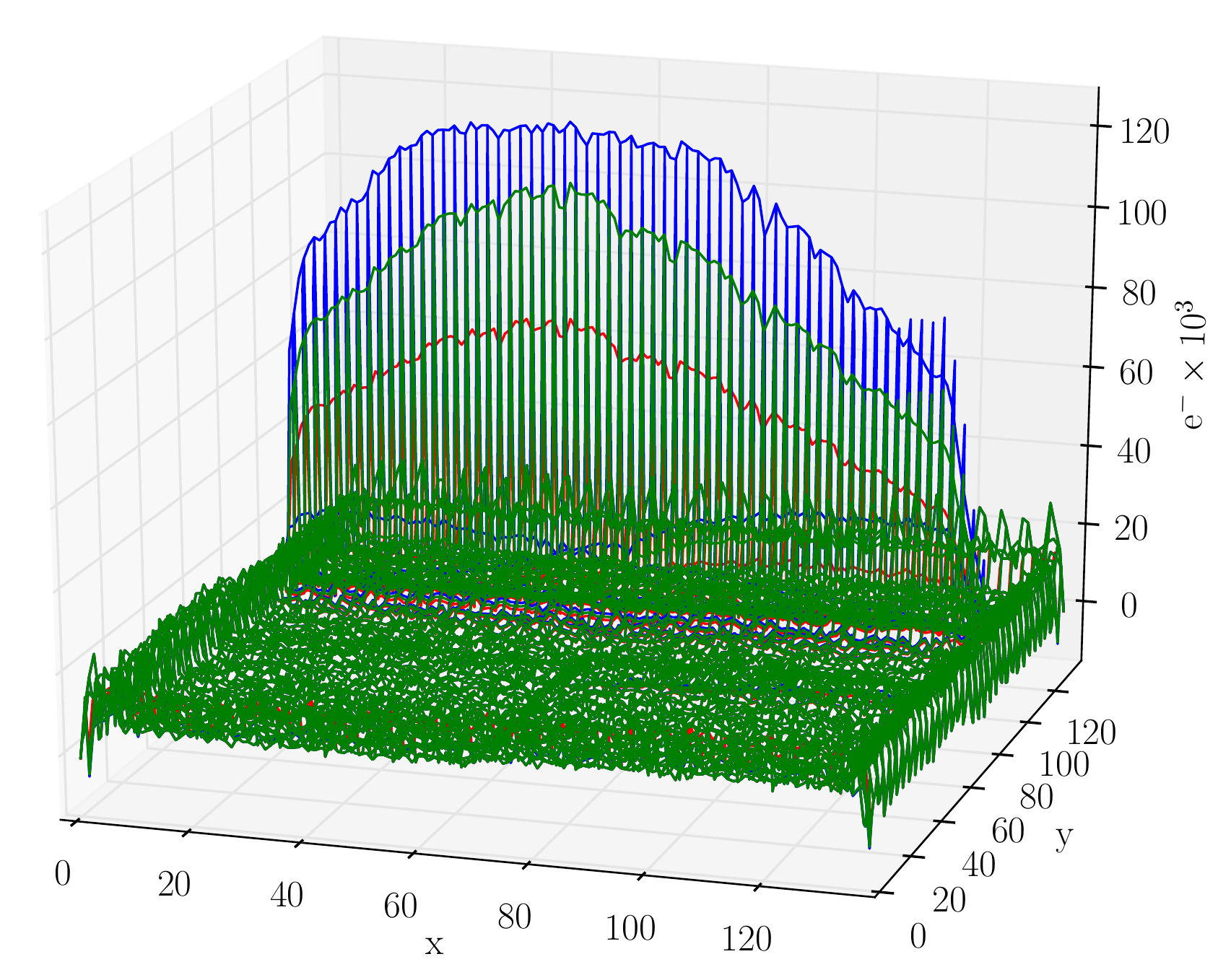}
\caption{Wireframe plot of the raw images that make up a typical exposure, with the zeroth (red), first (green) and second (blue) reads shown. The bias levels imposed during each detector reset have been subtracted, therefore showing the raw electron counts per pixel that contribute to non-linearity. The non-linearity limit of  $\gtrsim 78\,000$\,e$^-$/pixel is surpassed by a significant region in the centre of the first readout, and most of the second readout, therefore these regions cannot provide useful transit light curves.}
\label{fig:WFC3_wireframe}
\end{figure}

The data were reduced using the latest version of the {\sc  calwf3} data-pipeline ({\sc  v2.3}). This removes bias-drifts, subtracts the zeroth readout from each subsequent non-destructive read, subtracts dark images, and corrects each pixel for non-linearity. As the WFC3 has no shutter, as many as $\sim$\,60\,000\,e$^-$/pixel were recorded in the zeroth readout image. These are accumulated between the final reset of the detector before an exposure (the detector is continuously reset when not recording photons or being read out) and the time of zeroth read, and may also be influenced by persistence effects. The detector exhibits significant non-linear behaviour, which is correctable at low to intermediate levels, and takes the zeroth read counts into account. At high levels, $\gtrsim 78\,000$\,e$^-$/pixel, the response becomes highly non-linear and not correctable \citep{IHB_WFC3}\footnote{This is defined as where the detector counts deviate by more than 5\% from linearity.}. Unfortunately, for both the first and second readouts, and for both visits, a strip of pixels is saturated along the central region of the spectra. Fig.~\ref{fig:WFC3_wireframe} shows a plot of the raw images for one spectra, with the zeroth, first and second reads shown, after subtraction of the bias levels imposed during each detector reset\footnote{The bias level does not result from physical charges in each pixel therefore does not contribute to the detector non-linearity.}. It therefore shows the raw electron counts per pixel, that contribute to non-linearity. The non-linearity limit of  $\gtrsim 78\,000$\,e$^-$/pixel is surpassed by a significant region in the centre of the first readout, and most of the second readout, therefore these regions cannot provide useful transit light curves. We tried to extract useful data from the zeroth read spectra, but they resulted in unphysical transit depths, presumably related to persistence effects in the detector. Subtracting the zeroth read is important to remove this effect from the subsequent reads. The pipeline does not apply the non-linearity correction to the zeroth read prior to this. This is a potential source of error in using the current pipeline, but the recorded counts in the zeroth read are within the linear regime (at least in the regions of the spectrum we finally use), so we assume this effect is negligible on our final light curves. The pipeline usually only applies the non-linearity correction to non-saturated pixels; nonetheless we forced the pipeline to apply the correction to {\it all} pixels, and proceeded to extract spectra for both first and second readouts.

\begin{figure}
\includegraphics[width=85mm]{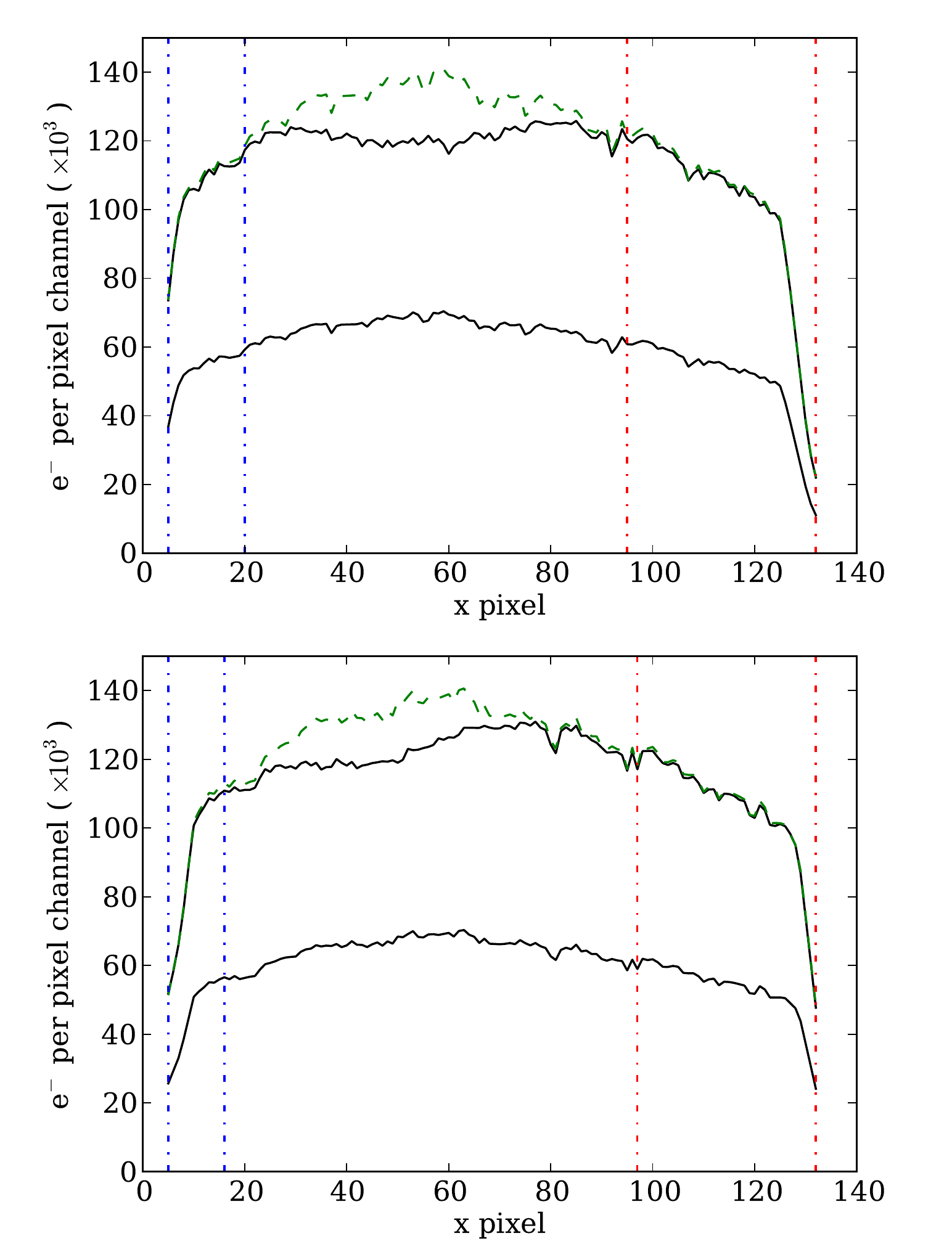}
\caption{Example spectra extracted for one image for the first (top) and second (bottom) visits, showing spectra for both the first and the second readouts. The dashed green line is the first readout scaled to the exposure time of the second. Clearly the spectra of the second readout are saturated, but much of the wavelength range in both the first and second readout is non-linear and not correctable to a scientifically useful level (see text). We therefore only extracted light curves from either end of the wavelength range. The blue and red dashed-dotted lines show the extraction regions used for each visit. Note that the zeroth read is subtracted prior to extracting these spectra, and they are summed along the y-axis, hence do not correspond in a simple way to the detector saturation limit.}
\label{fig:WFC3_spectra}
\end{figure}

To extract the spectra, we used a technique similar to that used in \citet{Gibson_2011} for \HST/NICMOS spectroscopy, using a custom pipeline written in {\sc python}. The spectral trace was first determined for each image by fitting a Gaussian profile in the cross-dispersion axis (y) for each of 128 columns along the dispersion direction (x). A straight line was fitted to the centres of the profiles to define the spectral trace. A sum of 10 pixels was taken along the cross-dispersion direction centred on the spectral trace to determine the flux for each pixel channel, after subtracting a background value. The extraction width was chosen to minimise the out-of-transit RMS after corrections for instrumental systematics were applied (see Sect.~\ref{sect:analysis}). The final light curves were not particularly sensitive to varying the extraction width between \s7\,--\,15 pixels. The background was determined from a region on the same pixel column above the spectrum, but we also tested a global background correction. The final light curves were not  sensitive to the choice of background region, given the low background counts ($\simeq$35\,e$^-$/pixel) and lack of spatial variation. Fig.~\ref{fig:WFC3_spectra} shows spectra extracted for both runs and for both the first and second readouts. The dashed line is first order spectra scaled to match the exposure time of the second readout. The saturated region is obvious in the second readout, but over much of the wavelength range in both the first and second readout the flux is non-linear.

Transits for each pixel channel were then constructed from time series of the spectra. The transits suffer from instrumental systematic noise, which is typical of \HST\ observations of transit light curves (\eg ACS: \citealt{Pont_2008}, NICMOS: \citealt{Swain_2008}, \citealt{Gibson_2011}, STIS: \citealt{Sing_2011}). These may arise due to thermal breathing and temperature variations, plus pointing drifts which aggravate pixel-to-pixel sensitivity variations. Transits in the saturated regions of the spectra predictably showed shallower than expected transit depths (often unphysical). We explored various methods to extract useful information from the saturated regions of the spectra. Indeed, this is why we chose to use a custom pipeline rather than standard extraction tools, as it allows much more control over the data reduction process, and the benefits of more sophisticated methods such as optimal extraction over aperture extraction are marginal for high S/N data. One such method was to simply mask the saturated pixels (the same pixels in every image), and sum over the remaining pixels. As the FWHM in the cross-dispersion direction is rather narrow ($\sim2$ pixels), the saturated pixel contained much of the signal. More importantly, the systematics resultant from pointing drifts were greatly enhanced when the central pixel is masked, as small variations in position cause a large change in the response. After further tests, reluctantly we concluded that useful information could not be obtained where the central pixel was saturated, and decided to extract two light curves at either end of the wavelength range in the first readout spectra, within the unsaturated regions. 

\begin{figure*}
\includegraphics[width=165mm]{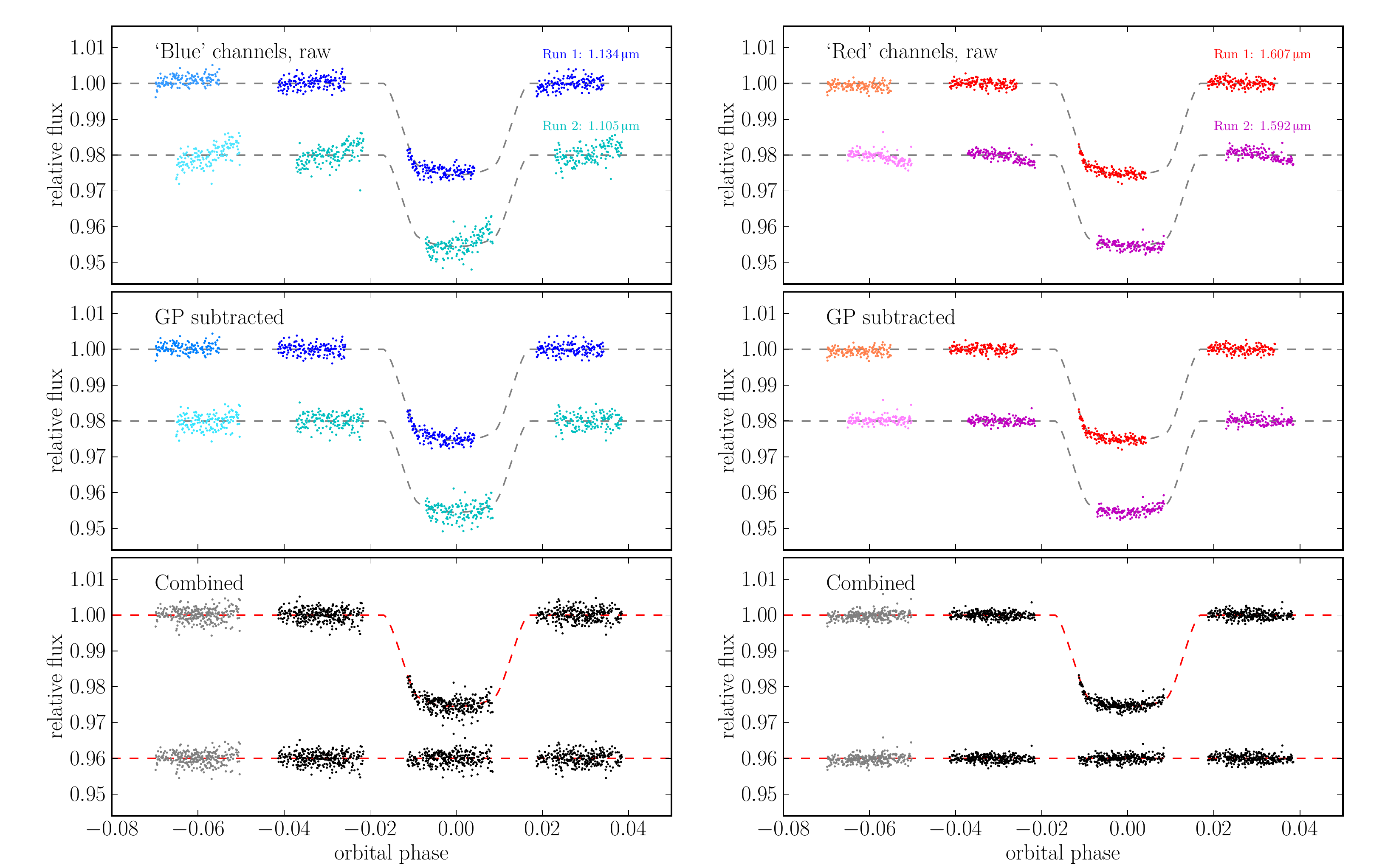}
\caption{Light curves for the blue (left) and red (right) channels. Top: Raw extracted light curves, showing systematics that need accounting for when inferring the planet-to-star radius ratio. Middle: `Cleaned' light curves after subtracting the Gaussian process systematics model. Bottom: Combined light curves from both visits. The cleaned light curves are shown for illustrative purposes only; the planet-to-star radius ratio is inferred simultaneously with the GP systematics model. The first orbit was excluded for the inference, but the projected GP model is subtracted nonetheless.}
\label{fig:WFC3_lightcurves}
\end{figure*}

For the first visit, we summed the first 15 pixel channels, and final 37 pixel channels to form what we will refer to as `blue' and `red' channels, respectively. Similarly, for the second visit we extracted the first 11 and final 35 pixel channels into the blue and red channels. These extraction regions are marked in Fig.~\ref{fig:WFC3_spectra}. Fig.~\ref{fig:WFC3_lightcurves} shows the extracted blue and red channel light curves for the first and second visits. The wavelength dispersion of the WFC3/IR G141 grism is position dependent, and the position of HD 189733 in the direct image was used to calibrate the wavelengths. The blue and red channels in the first visit correspond to wavelength ranges of $1.099 - 1.168$\,\micron\ and $1.521 - 1.693$\,\micron, respectively, and $1.082 - 1.128$\,\micron\ and $1.514 - 1.671$\,\micron, respectively, for the second transit.

Instrumental systematics are evident in the light curves, and must be accounted for to measure the planet-to-star radius ratio. We therefore extracted auxiliary parameters from the data in a similar way to \citet{Gibson_2011} in order to explore systematics models. These were the shift in x-position ($\Delta X$), y-position ($\Delta Y$), and average width ($W$) and angle ($\psi$) of the spectral trace. $\Delta X$ was determined from cross-correlation of the extracted spectra in each image (first readout), $\Delta Y$ and $\psi$ from the straight line fits to the spectral trace, and $W$ from the average FWHM of the Gaussian profiles fitted to each extraction column in the cross-dispersion direction. The phase ($\theta_{HST}$) of \HST\ was also calculated for each time. Unfortunately no useful temperature information was available in the image headers. \citet{Berta_2011} provide a detailed description of WFC3 systematics for transmission spectroscopy, which we do not attempt to reproduce here. However, we note that the form of the systematics in our light curves are slightly different given the different observing strategies used; most noteworthy we do not see a prominent `ramp' effect. This is likely because we do not require buffer dumps during orbits, which leads to regular cadence, therefore stable persistence effects for each exposure.

\section{Analysis}
\label{sect:analysis}

\subsection{Gaussian process model}

The light curves were modelled using the GP model described in \citet{Gibson_2011b}, which we briefly recap here. Each light curve is modelled as a GP with a transit mean function:
\[
f(t, \bmath{x}) \sim \mathcal{GP} \left (T(t,\bphi) , \mathbf\Sigma(\bmath x, \btheta)  \right),
\]
where $f$ is the flux, $t$ the time, $\bmath x$ the vector containing $K$ auxiliary measurements, $T$ is the transit (mean) function with transit (hyper-)parameter vector $\bphi$ (including the planet-to-star radius ratio, $\rho$), modelled using the analytic equations of \citet{Mandel_Agol_2002}, and $\mathbf\Sigma$ is the covariance matrix with hyperparameter vector $\btheta$. The covariance between two points $\mathbf{x}_n$ and $\mathbf{x}_m$ is given by the squared exponential kernel function:
\[
\mathbf{k}(\mathbf{x}_n,\mathbf{x}_m) = \xi \exp\left(-\sum_{i=1}^K\eta_i(x_{n,i} - x_{m,i})^2\right) + \delta_{nm} \sigma^2,
\]
where $\xi$ is a hyperparameter that specifies the maximum covariance, $\eta_i$ are the inverse scale hyperparameters for each auxiliary measurement vector, $\sigma^2$ is the variance due to white noise, and $\delta_{nm}$ is the Kronecker delta. The kernel accounts for the instrumental systematics, and this one describes a smooth function of the input parameters, with the addition of white noise. In other words, the instrumental systematics model assumes a similar value when all {\it relevant} auxiliary measurements are close, with the `closeness' specified by the inverse length scales. Similarly to other \HST\ analyses, we excluded orbit 1 from the fits. This has the added advantage of significantly speeding up inversion of the covariance matrix and therefore inference of the planet-to-star radius ratio for each light curve.

The definition of a GP states that the likelihood is multivariate Gaussian. We multiply the likelihood by gamma hyperpriors of shape parameter unity for $\xi$ and the $\eta_i$ hyperparameters to form the posterior distribution (implying uniform, improper priors for the remaining parameters). The scale length of the hyperpriors were set to large values so they have little influence over the inference, other than to ensure the hyperparameters remain positive. The posterior distribution of the planet-to-star radius ratio ($\rho$) is found by marginalising over all the other transit parameters and covariance hyperparameters using Markov-Chain Monte-Carlo (MCMC), where the fixed transit parameters\footnote{We fix the scale length $a/R_\star$ and inclination $i$; these parameters are correlated with the planet-to-star radius ratio, but in transmission spectroscopy we are interested in the posterior distribution of $\rho$ conditioned on these parameters, \ie the relative values for $\rho$.} are set to the values of \citet{Pont_2007,Pont_2008} except the ephemeris which is calculated from the \citet{Agol_2010} values\footnote{The more up to date physical parameters of \citet{Agol_2010} were not used here to enable direct comparison with NICMOS and ACS data.}. The light curve function is multiplied by a normalisation (baseline) function during fitting, in this case a linear function of time, whose parameters we also marginalise over in the MCMC. In all, we fit for the planet-to-star radius ratio, the parameters of the normalisation function, and all the GP hyperparameters including the white noise. For each light curve we ran four chains and tested for convergence using the Gelman \& Rubin (GR) statistic \citep{GelmanRubin_1992}. For more details on the GP model and MCMC routine, see \citet{Gibson_2011b}. The limb-darkening parameters were calculated for each light curve using the methods described in \citet{Sing_2010} for both quadratic and non-linear limb darkening laws. These were calculated using the 1D stellar atmosphere models rather than the 3D model atmospheres used in \citet{Sing_2011}. This was to enable easy comparison with ACS and NICMOS transmission spectroscopy, but future studies will use the more reliable 3D limb darkening parameters. Inference of the planet-to-star radius ratio was performed for both laws holding the limb darkening parameters fixed. 

\begin{figure}
\includegraphics[width=85mm]{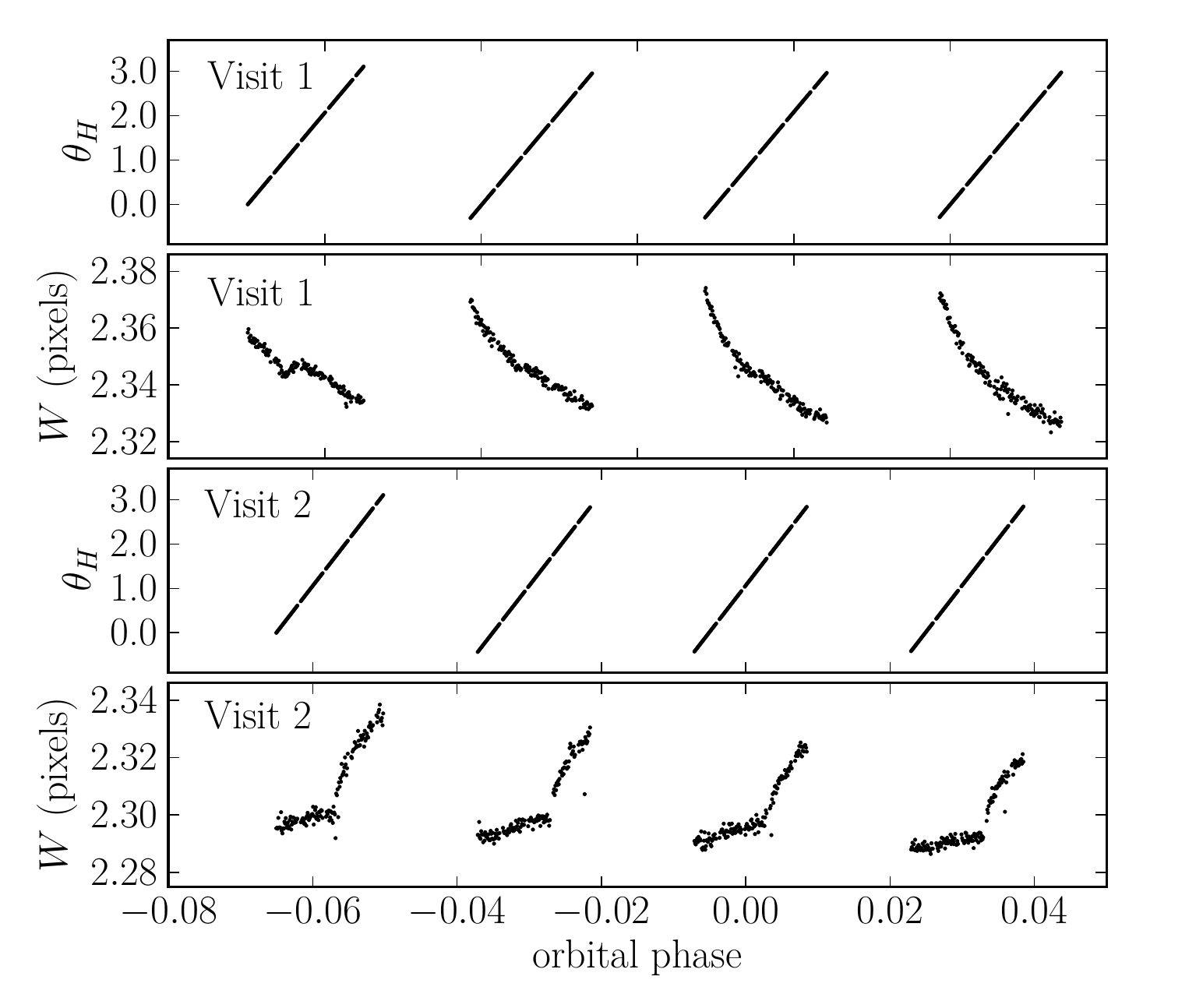}
\caption{Plots of $\theta_{HST}$ and $W$ for both visits as a function of orbital phase of HD 189733. These are the auxiliary inputs used as input vectors for the final GP model of the instrumental systematics.}
\label{fig:WFC3_decorr_params}
\end{figure}

We decided to take a different approach than for the NICMOS data; rather than use all of the auxiliary information, we were careful to use only the inputs that are needed to model the shape of the systematics per orbit, {\it but importantly do not change the flux offsets between orbits.} Therefore we first tried using only the orbital phase of \HST\ ($\theta_{HST}$) as input to the GP. This did not provide a satisfactory correction for all orbits, particularly for orbits where a discontinuity is evident mid way through each orbit in the light curves\footnote{The discontinuity is caused by \HST\ crossing the Earth's shadow.}. We therefore included the cross-dispersion width ($W$) of the spectrum as an additional GP input. Plots of $\theta_{HST}$ and $W$ for both visits are shown in Fig.~\ref{fig:WFC3_decorr_params}. These were normalised to zero mean and variance of 1 prior to their use as GP inputs to ensure the hyperpriors provide similar constraints. This GP model provided a satisfactory correction for the systematics, and the two visits gave consistent results. The MCMC chains were well behaved in this case, and the GR statistic was within 1\% of unity. These are used as our final values for $\rho$. We also tested this model including orbit 1 in the fitting procedure, which did not significantly change the results, nor did using (GP) smoothed versions of the $W$ auxiliary parameter.  The light curves produced after subtracting the GP systematics models are shown in Fig.~\ref{fig:WFC3_lightcurves}.

We further tested the GP model using $\Delta X$, $\Delta Y$ and $\psi$ as additional GP inputs. The MCMC did not converge in this case, or give consistent results for the two runs.
These additional inputs show similar structure to the $W$ input but are much noisier. This causes degeneracy in the GP hyperparameters thus demanding much longer MCMC chains to achieve convergence. However, we also ran tests using GP smoothed input parameters \citep[see][]{Gibson_2011b}, and the MCMC now converged, showing that the level of noise on the inputs can also significantly affect convergence. This led to consistent values for $\rho$ when compared to using only $\theta_{HST}$ and $W$ as inputs, although with larger uncertainties. Given that these additional inputs are particularly noisy, and more importantly that they show (small) offsets between orbits, we chose not to use these parameters in our final results, as this can cause flux offsets to be fitted by the GP. For $\psi$, the in-transit values are also outside the range covered by the out-of-transit data, hence requiring an extrapolation to the in-transit data. In theory, these issues might be addressed with appropriate prior widths in the gamma hyperpriors, but this becomes rather subjective with no robust way to set them. Of course, in an ideal Bayesian framework we should be able to include all the (noisy) inputs. More sophisticated tools to evaluate the posterior distribution and marginalise over the hyperparameters such as Bayesian Quadrature methods \citep[\eg][]{OHagan_1991} might enable this in the future.

Our approach of using $\theta_{HST}$ and $W$ as the only inputs is supported by two further pieces of evidence. The first is that the intrapixel sensitivity variations are much smaller than for NICMOS, with no measurable variation found during ground-testing for WFC3, compared with $\sim40\%$ for NICMOS \citep{IHB_WFC3}; therefore the variations in flux caused by shifts in $X$, $Y$ and the angle should have little or no impact. The second is that recent WFC3 analysis by \citet{Berta_2011} did not require modelling any flux offsets between orbits to fully account for the systematics. Indeed the `divide-oot' method used should provide a similar correction to the GP model using $\theta_H$ and $W$ as inputs, given that they show regular repeating structure for each orbit.

\subsection{Linear basis function model}

As an independent check we also tried using the `standard' method to model the systematics \citep[see \eg][]{Brown_2001b,Gilliland_2003,Pont_2007,Swain_2008}, \ie using the extracted auxiliary parameters as basis functions, and fitting for the coefficients using linear least squares (see \citealp {Gibson_2011,Gibson_2011b} for a detailed description of the method used here). Using $W$ and $\theta_{HST}$ as basis functions, and higher order squared terms, provided consistent results with the GP model (within 1\,$\sigma$), and the residual permutation method described in \citet{Gibson_2011} provided similar uncertainties. When we added $\Delta X$, $\Delta Y$ and $\psi$ as additional basis vectors the instrument model became unstable and the two visits provided different planet-to-star radius ratios for the blue and red transits, validating our use of $W$ and $\theta_{HST}$ as the only inputs to the GP.

\section{Results}
\label{sect:results}

\subsection{WFC3 light curve fits}

The inferred planet-to-star radius ratio for both quadratic and non-linear limb darkening laws are given in Tab.~\ref{tab:results}\footnote{We have simply given the wavelength range of the extraction region; more detailed analyses would need to take into account the spectral response, but is not warranted here.}, as both have been used in the literature for HD 189733. The uncertainties are calculated from the distribution of the MCMC chains, after removing the first 20\% of each, and are given by the limits which encompass 68.2\% of the probability distribution.

\begin{figure}
\includegraphics[width=85mm]{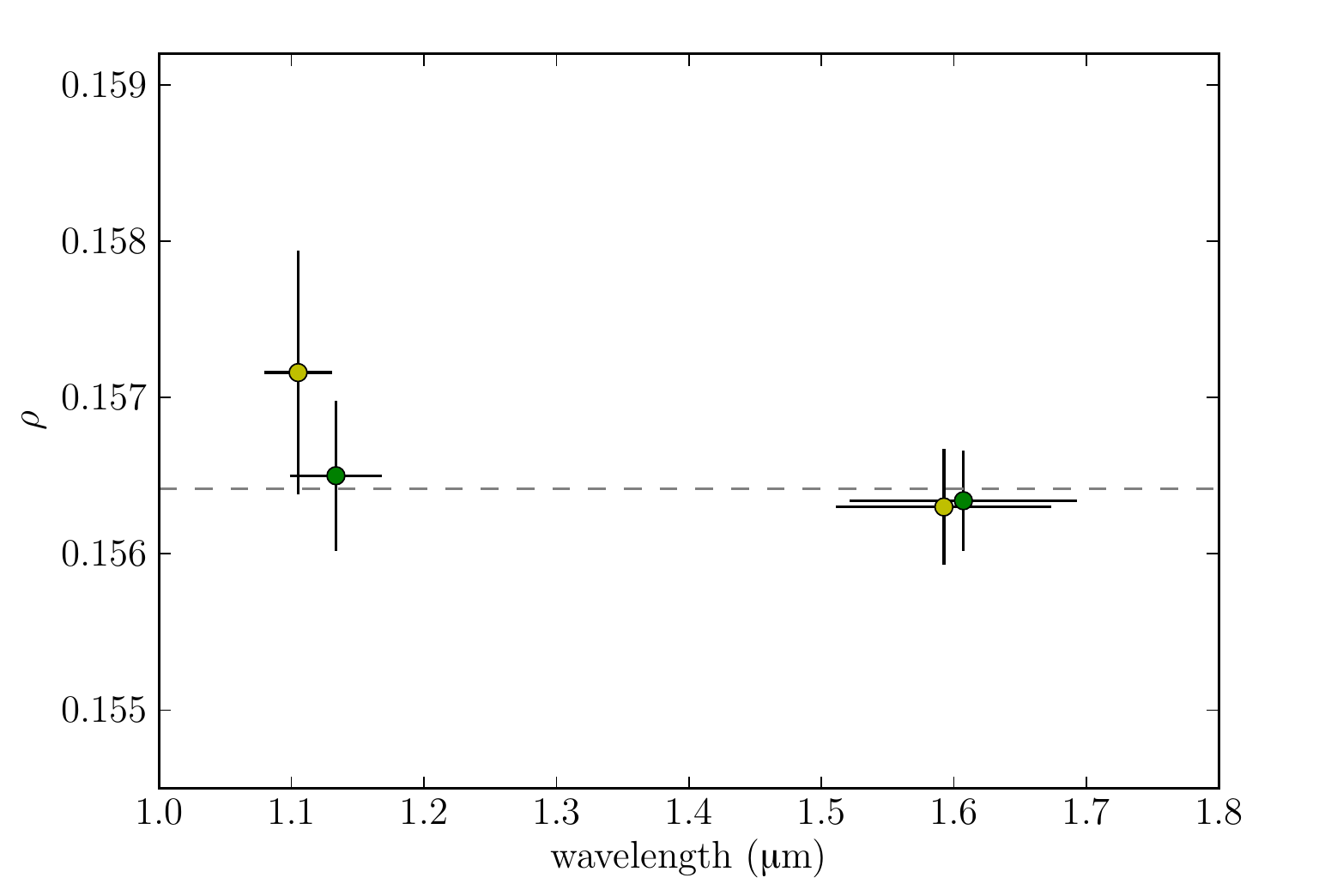}
\caption{Measured planet-to-star radius ratio of HD 189733b from WFC3 data using quadratic limb darkening prior to spot corrections. Visit 1 and 2 are shown by the green and yellow points, respectively. The dashed horizontal line marks the weighted average of all four points.}
\label{fig:WFC3_trans_spectra_solo}
\end{figure}

The planet-to-star radius ratio obtained with the quadratic limb darkening law are plotted in Fig.~\ref{fig:WFC3_trans_spectra_solo}, and the dashed line is a weighted mean of all four data points. Clearly no planet-to-star radius ratio variation is detected in the WFC3 data. A comparison with other data requires corrections for unocculted star spots, discussed in the following section.

\begin{table}
\caption{Planet-to-star radius ratio calculated from the GP inference, given for both limb darkening laws.}
\label{tab:results}
\begin{tabular}{lcc}
\hline
Wavelength ($\micron$) & $\rho~(\frac{R_p}{R_\star})$ & visit\\
\hline

\multicolumn{3}{l}{(quadratic limb darkening)}\\
~$1.099-1.168$ & $0.15650\pm0.00048$ & 1 \\
~$1.082-1.128$ & $0.15716\pm0.00078$ & 2 \\
~$1.521-1.693$ & $0.15634\pm0.00032$ & 1 \\
~$1.514-1.671$ & $0.15630\pm0.00037$ & 2 \\
\multicolumn{3}{l}{(non-linear limb darkening)}\\
~$1.099-1.168$ & $0.15650\pm0.00047$ & 1 \\
~$1.082-1.128$ & $0.15735\pm0.00076$ & 2 \\
~$1.521-1.693$ & $0.15644\pm0.00032$ & 1 \\
~$1.514-1.671$ & $0.15672\pm0.00043$ & 2 \\
\hline
\end{tabular}
\end{table}

\subsection{Spot corrections}

As HD 189733 is an active star, we need to correct the transmission spectrum for unocculted star spots to make a detailed comparison with data taken at different times and wavelengths. Here, we use the spot corrections as described in \citet{Sing_2011} in order to place the WFC3 data in context alongside other transmission spectra of HD 189733b. However, given the complexity of spot corrections, this warrants its own publication, and will form a large part of a future paper (Pont et al, in prep), where we will analyse the complete transmission spectra of HD 189733b from UV to IR. In addition it is important to model the transit light curves with consistent stellar parameters and limb darkening laws, which will also be addressed in this paper.

Correcting for unocculted spots requires long term monitoring of the stellar flux to obtain the variation in spot coverage between visits, and an estimate of the flux of the unspotted surface. These were obtained by fitting long-baseline ground-based coverage from the T10 0.8 m Automated Photoelectric Telescope (APT) at Fairborn Observatory in southern Arizona, detailed in \citet{Henry_2008} that covers all the \HST\ visits considered here. The photometry was fitted using GP regression with the quasi-periodic kernel given in \citet{Aigrain_2011}. The spot correction was then performed in the same way as described in \citet{Sing_2011}, assuming a spot temperature of 4250\,K measured from STIS and ACS transits. The spot corrections shifted the planet-to-star radius ratio for the blue and red light curves by $-0.0010$ and $-0.0006$, respectively, for visit one, and $-0.0009$ and $-0.0005$ for visit 2.

We note that no spot-crossing events are visible in our data. It is highly likely that spots were occulted given that the optical transits almost always show spot-crossing events \citep[\eg][]{Pont_2007,Sing_2011}, rather they are not visible in our data given the lower S/N and lower flux contrast between the spotted and unspotted stellar surface. However, these may still influence the extracted transmission spectrum.

\begin{figure*}
\includegraphics[width=155mm]{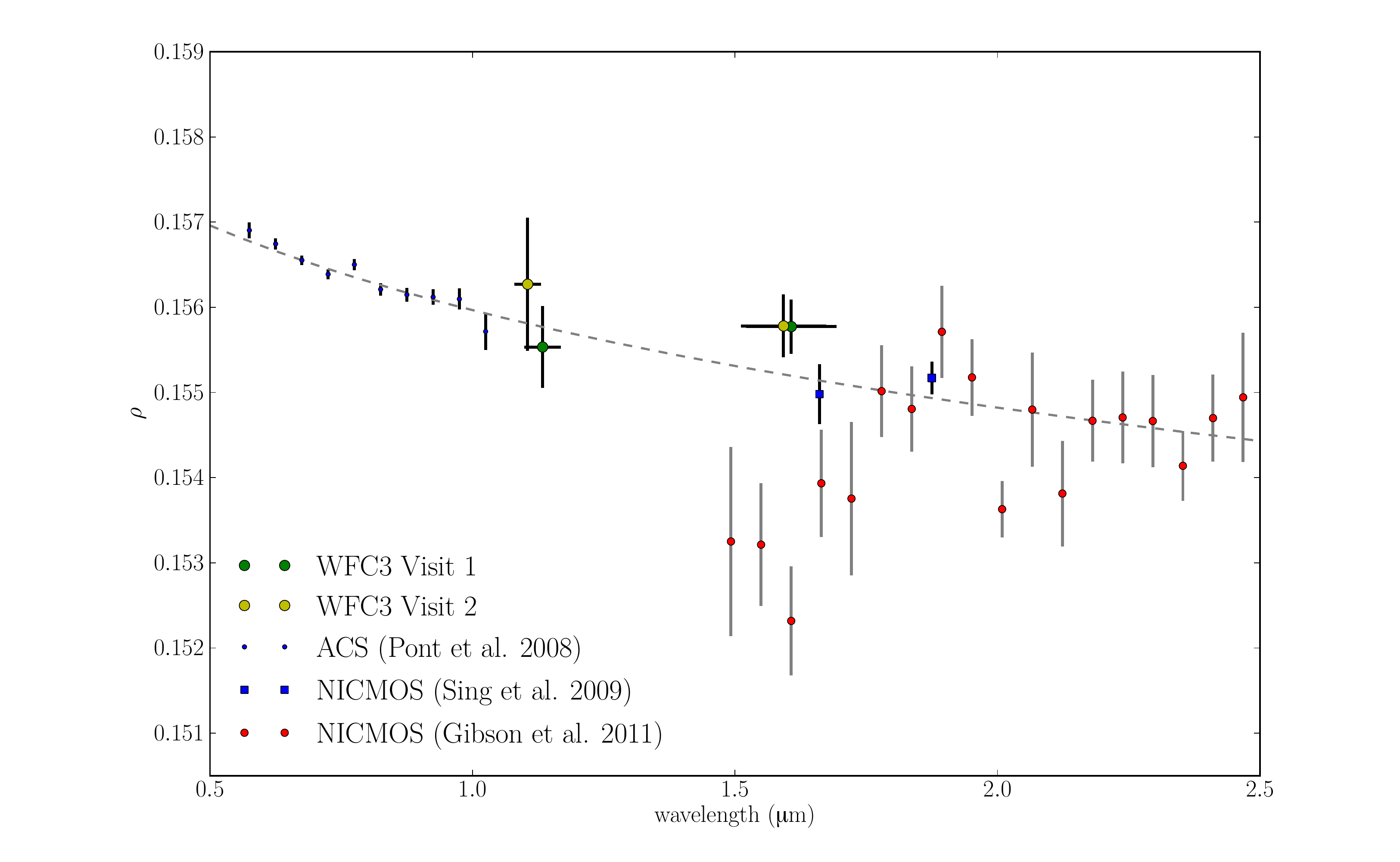}
\caption{Transmission spectra of HD 189733 showing the WFC3 data presented in this paper, the ACS data from \citet{Pont_2008}, the NICMOS photometry from \citet{Sing_2009}, and the reanalysis of the NICMOS transmission spectroscopy from \citet{Gibson_2011b} after correction for unocculted spots (see text). The dashed line is the Rayleigh scattering haze function given in \citet{Lecavelier_2008a}. The WFC3 blue and red light curves show consistent depth, providing no evidence for a `drop-off' in the haze observed at optical wavelengths. The WFC3 data are also consistent with the Rayleigh scattering haze extrapolated into the NIR; however given the difficulties extracting the light curves and performing the spot corrections we hesitate to draw firm conclusions from the WFC3 data.}
\label{fig:WFC3_trans_spectra}
\end{figure*}

Fig.~\ref{fig:WFC3_trans_spectra} plots the corrected WFC3 data alongside the ACS data from \citet{Pont_2008}, the NICMOS photometry from \citet{Sing_2009}, and the reanalysis of the NICMOS transmission spectroscopy from \citet{Gibson_2011b}. The ACS data were already corrected for spots using a similar technique and were taken directly from the paper; the NICMOS spectroscopy and photometry were corrected in the same way as the WFC3 data. The WFC3 data are largely consistent with ACS and NICMOS photometric data, but disagrees with the red end of the NICMOS spectroscopy. This tentatively suggests that the transmission spectrum is dominated by the haze into the NIR, and the feature at \s1.6\micron\ may indeed be of instrumental origin \citep{Gibson_2011b}. However, given the difficulty in performing spot corrections (particularly with the large baseline between the data sets), and in extracting the WFC3 data, it is not wise to rule out the feature at \s1.6\micron\ from these data alone.

\section{Discussion and Conclusion}
\label{sect:conclusions}

We have presented \HST /WFC3 spectroscopic observations of the transiting system HD 189733. The aim of this study was to bridge the gap between the ACS optical data of \citet{Pont_2008}, near-infrared transmission spectra with NICMOS \citep{Swain_2008,Gibson_2011b}, and NICMOS photometry of \citet{Sing_2009}. Unfortunately, the WFC3 spectra were saturated for much of the central region, and we could only extract useful light curves at either extreme of the G141 grism, corresponding to central wavelengths of $\sim$1.134\,\micron\ and 1.607\,\micron\ for visit 1 and $\sim$1.105\,\micron\ and 1.592 \,\micron\ for visit 2. The red and blue light curves show consistent transit depths for both runs when using either quadratic or non-linear limb darkening laws. We have not detected a drop in the planet-to-star radius ratio one would expect if the optical haze does indeed become transparent between $\sim$1.1 and 1.4\,\micron, as has been suggested to explain the different behaviour observed at optical and near-infrared wavelengths.

Correction for unocculted star spots shows the WFC3 data are consistent with the ACS data of \citet{Pont_2008} and the NICMOS photometry of \citet{Sing_2009}, but disagree with the red end of the NICMOS transmission spectroscopy of \citet{Swain_2008} and the reanalysis of \citet{Gibson_2011b}. A more detailed analysis of the complete transmission spectrum of HD 189733b is in preparation, with more detailed spot corrections, consistent limb darkening laws and stellar parameters. Nonetheless, this work tentatively suggests that Rayleigh scattering from the high-altitude haze dominates the transmission spectrum up to and including near-infrared wavelengths. We note that both the WFC3 data and the NICMOS photometry of \citet{Sing_2009} have shown consistent transit depths, and have failed to produce the deep feature at \s1.6\,\micron\ reported in \citet{Swain_2008}. However, whilst WFC3 appears better behaved than NICMOS transmission spectroscopy, given that the data are near the non-linear range of the detector, and the difficulty in performing spot corrections, more evidence is required to conclusively resolve the issue and provide continuous coverage from optical to near-infrared wavelengths. Future observations of HD 189733 in `driftscan' mode should help provide this evidence.

\section*{Acknowledgments}

All of the data presented in this paper were obtained from the Multimission Archive at the Space Telescope Science Institute (MAST). STScI is operated by the Association of Universities for Research in Astronomy, Inc., under NASA contract NAS5-26555. Support for MAST for non-HST data is provided by the NASA Office of Space Science via grant NNX09AF08G and by other grants and contracts. N. P. G and S. A. acknowledge support from STFC grant ST/G002266/2. We are extremely grateful for the support provided the WFC3 instrument team, in particular H. Bushouse, and discussions with P. McCullough. Finally, we thank the referee, R. Gilliland, for his careful reading of the manuscript and helpful suggestions.

\bibliography{../MyBibliography} 
\bibliographystyle{mn2e_astronat} 

\label{lastpage}

\end{document}